\documentstyle[preprint,aps]{revtex}

\begin{document}
\draft
\title{
Exact Critical Properties of the Multi-Component \\
Interacting Fermion Model with Boundaries
}
\vskip1cm
\author{Satoshi Fujimoto}
\address{
Department of Physics,
Kyoto University, Kyoto 606, Japan
}
\author{Norio Kawakami}
\address{
Department of Applied Physics,
and Department of Material and Life Science, \\
Osaka University, Suita, Osaka 565, Japan
}
\maketitle
\begin{abstract}
Exact critical properties of the one-dimensional SU($N$)
interacting fermion model with open boundaries are studied
by using the Bethe ansatz method.
We derive the surface critical exponents of various correlation
functions using boundary conformal field theory.
They are classified into two types,
i.e. the exponents for the chiral SU($N$)
Tomonaga-Luttinger liquid and those related to the orthogonality
catastrophe. We discuss a possible application of the results
to the photoemission (absorption) in the edge
state of the fractional quantum Hall effect.
\end{abstract}
\vspace{1 cm}
\narrowtext
\section{Introduction}

One-dimensional (1D) quantum many-body
systems with open boundaries
have attracted much interest recently in connection with various
problems in condensed matter physics such as the Kondo problem,
the tunneling in a quantum wire and the edge state
of the fractional quantum Hall effect(FQHE).
A number of exactly solvable models with open boundaries
have been known so far, {\it e.g.} the XXZ Heisenberg
model\cite{bxxz},
the interacting boson model\cite{gaud},
the interacting fermion model\cite{woy},
the Hubbard model\cite{shu}, and the $1/r^2$
quantum model\cite{yam}, etc. Quantum impurity models
such as the Kondo model and the Anderson model
also have a deep connection with the boundary
problem\cite{alkondo,fky2}.
All the above systems exhibit the
surface critical behavior near the boundary.
The corresponding surface exponents,
which should be different from the bulk ones, can
be obtained by using the finite-size
scaling in boundary conformal field theory\cite{cardy}.

In this paper, we obtain the exact surface critical exponents
for the 1D SU($N$) interacting fermion model with
open boundaries
by combining the Bethe ansatz solution and boundary conformal
field theory. By examining boundary conditions carefully, we
classify the surface exponents into
two categories, i.e. the exponents for the chiral
Tomonaga-Luttinger liquid and those related
to the orthogonality catastrophe.  The latter is shown to be
related to the X-ray problem in 1D chiral systems.
We apply our results to the edge states of the  FQHE,
and predict some expected behaviors for the photoemission
(absorption) singularity.
Some of the results obtained in this paper
were previously conjectured in \cite{fky2}.

The plan of the paper is as follows.
In sec. II, we introduce the continuum fermion model with
SU($N$) spin symmetry in open
boundary conditions, and give the Bethe ansatz equations.
In sec. III, based on boundary conformal field theory,
we derive critical exponents  for various
correlation functions from the finite-size
spectrum computed by the Bethe ansatz solution.
In sec.IV, we discuss a possible application
of the present results to the X-ray photoemission (absorption)
problem in 1D chiral
electron systems. A brief summary is given in  sec.V.
\section{Model and Bethe ansatz solution}

We consider the interacting fermion model with
SU($N$) spin symmetry in open boundary conditions.
The mutual electron interaction is of $\delta$-function type, and
boundary potentials are introduced
at both ends of the open chain. The Hamiltonian is thus given by
\begin{equation}
H=-\sum_{m=1}^{N}\sum_{i=1}^{N_{m}}
\frac{\partial^2}{\partial x^2_{i,m}}
+2c\sum_{i<j,m,n}\delta(x_{i,m}-x_{j,n})
-\sum_{m=1}^{N}\sum_{i=1}^{N_{m}}(\gamma_0\delta(x_{i,m})
+\gamma_L\delta(x_{i,m}-L)), \quad (c>0)
\label{hamil}
\end{equation}
where $N_{m}$ is the number of electrons
with spin index $m$ ($=1,2, \cdots, N$) of
SU($N$) symmetry and $L$ is the system size. The last two terms
represent boundary potentials with
coupling constants $\gamma_{0(L)}$.
We note that this type of the SU($N$)
fermion model with $\delta$-function interaction  was
first solved  by Sutherland under the periodic
boundary condition many years
ago\cite{suth}.  We now wish to solve the above model
(\ref{hamil}) for the open chain
with boundary potentials  $\gamma_{0(L)}$.
This can be preformed by generalizing
Gaudin's method developed for the boson model
with open boundary conditions\cite{gaud}.  The main idea is that
one can treat the open boundary problem more easily by introducing
fictitious {\it mirror-image particles} with respect to the
boundary. We then diagonalize the Hamiltonian
(\ref{hamil}) following standard Bethe ansatz techniques
developed for periodic systems\cite{suth},
and end up with the basic algebraic equations
for rapidities $k_j$ and $\Lambda_{\alpha}^{(l)}$,
\begin{eqnarray}
2Lk_j+\varphi_0(k_j)+\varphi_L(k_j)=2\pi I_j
&-&\sum_{\beta=1}^{M_1}\biggl(2\tan^{-1}\biggl[\frac{2(k_j
-\Lambda_{\beta}^{(1)})}{c}\biggr]
+2\tan^{-1}\biggl[
\frac{2(k_j+\Lambda_{\beta}^{(1)})}{c}\biggr]\biggr),
\label{eqn:bethe11}\\
\sum_{\beta=1}^{M_{l-1}}\biggl(
2\tan^{-1}\biggl[\frac{2(\Lambda_{\alpha}^{(l)}
-\Lambda_{\beta}^{(l-1)})}{c}\biggr]
&+&2\tan^{-1}\biggl[\frac{2(\Lambda_{\alpha}^{(l)}
+\Lambda_{\beta}^{(l-1)})}{c}\biggr]\biggr)=
2\pi J_{\alpha}^{(l)} \nonumber \\
+\sum_{\beta=1,\beta \neq\alpha}^{M_{l}}\biggl(
2\tan^{-1}\biggl[\frac{\Lambda_{\alpha}^{(l)}
-\Lambda_{\beta}^{(l)}}{c}\biggr]
&+&2\tan^{-1}\biggl[\frac{\Lambda_{\alpha}^{(l)}
+\Lambda_{\beta}^{(l)}}{c}\biggr]\biggr) \nonumber \\
-\sum_{\beta=1}^{M_{l+1}}\biggl(
2\tan^{-1}\biggl[\frac{2(\Lambda_{\alpha}^{(l)}
-\Lambda_{\beta}^{(l+1)})}{c}\biggr]
&+&2\tan^{-1}\biggl[\frac{2(\Lambda_{\alpha}^{(l)}
+\Lambda_{\beta}^{(l+1)})}{c}\biggr]\biggr),  \nonumber \\
 & &  1\leq l \leq N-1, \quad \alpha=1,2,...,M_l,
\label{eqn:bethe12}
\end{eqnarray}
with  $M_{N}\equiv 0$, $M_l=\sum_{\alpha=l+1}^{N}N_{\alpha}$
($0\leq l \leq N-1$) and $\Lambda^{(0)}_j\equiv k_j$, where
$\varphi_p(k)=-2\tan^{-1}\frac{k}{\gamma_p}$, ($p=0, L$)
are the phase shifts due to the boundary scattering.
Quantum numbers $I_j$ and $J_{\alpha}^{(l)}$
are positive integers (or half-odd integers) which classify
the elementary excitations. The total energy is simply given by
\begin{equation}
E=\sum_{j=1}^{N} k^2_{j}.
\label{eqn:en}
\end{equation}
Note that the above equations contain the terms with
arguments like $\Lambda_{\alpha}^{l}+\Lambda_{\beta}^{l}$,
reflecting the introduction of image particles\cite{gaud,woy,shu},
in contrast to the ordinary Bethe ansatz equations for
the periodic case.
These equations for the SU(2) case were obtained firstly by
Woynarovich\cite{woy}. Defining new variables,
$\Lambda_{-\alpha}^{(l)}=-\Lambda^{(l)}_{\alpha}$,
we can rewrite the above equations into more tractable form,
\begin{eqnarray}
2Lk_j+\varphi_0(k_j)+\varphi_L(k_j)=2\pi I_j
&+&2\tan^{-1}\biggl(\frac{2k_j}{c}\biggr)
-\sum_{\beta=-M_1}^{M_1}2\tan^{-1}\biggl[\frac{2(k_j
-\Lambda_{\beta}^{(1)})}{c}\biggr], \label{eqn:bethe21}\\
\sum_{\beta=-M_{l-1}}^{M_{l-1}}
2\tan^{-1}\biggl[\frac{2(\Lambda_{\alpha}^{(l)}
-\Lambda_{\beta}^{(l-1)})}{c}\biggr]&=&
2\pi J_{\alpha}-2\tan^{-1}
\biggl(\frac{\Lambda_{\alpha}^{(l)}}{c}\biggr)
+2\tan^{-1}\biggl(\frac{2\Lambda_{\alpha}^{(l)}}{c}\biggr)
\nonumber \\
+\sum_{\beta=-M_{l}}^{M_{l}}
2\tan^{-1}\biggl[\frac{\Lambda_{\alpha}^{(l)}
-\Lambda_{\beta}^{(l)}}{c}\biggr]
&-&\sum_{\beta=-M_{l+1}}^{M_{l+1}}
2\tan^{-1}\biggl[\frac{2(\Lambda_{\alpha}^{(l)}
-\Lambda_{\beta}^{(l+1)})}{c}\biggr],  \quad 1\leq l \leq N-2,
\label{eqn:bethe22}\\
\sum_{\beta=-M_{N-2}}^{M_{N-2}}
2\tan^{-1}\biggl[\frac{2(\Lambda_{\alpha}^{(N-1)}
-\Lambda_{\beta}^{(N-2)})}{c}\biggr]&=&
2\pi J_{\alpha}-2\tan^{-1}
\biggl(\frac{\Lambda_{\alpha}^{(N-1)}}{c}\biggr)
\nonumber \\
&+&\sum_{\beta=-M_{N-1}}^{M_{N-1}}
2\tan^{-1}\biggl[\frac{\Lambda_{\alpha}^{(N-1)}
-\Lambda_{\beta}^{(N-1)}}{c}\biggr].
\label{eqn:bethe23}
\end{eqnarray}
After this transformation, the structure of the equations
formally resembles to that for the periodic case \cite{suth}.
This fact makes the following analysis much easier .

In the following, we will be concerned with the case of
repulsive boundary potentials ($\gamma_p<0$)
for simplicity. All the rapidities then turn out to be real
in this case. Taking the thermodynamic limit, we now recast the
algebraic equations (\ref{eqn:bethe21})
$\sim$(\ref{eqn:bethe23}) into the integral equations
for the density functions of rapidities,
\begin{eqnarray}
2\pi \rho (k)&=&2+\frac{1}{L}(\varphi_0' (k)+\varphi_L' (k))
-\frac{1}{L}\frac{c}{(c/2)^2+k^2} \nonumber \\
& &+\int^{\Lambda^{(1)}_{+}}_{\Lambda^{(1)}_{-}}d\Lambda
\frac{c}{(c/2)^2+(k-\Lambda)^2}\sigma^{(1)}(\Lambda),
\label{eqn:density1}\\
2\pi\sigma^{(l)}(\Lambda^{(l)})&=&
\frac{1}{L}\frac{2c}{c^2+(\Lambda^{(l)})^2}
-\frac{1}{L}\frac{c}{(c/2)^2+(\Lambda^{(l)})^2}
-\int^{\Lambda^{(l)}_{+}}_{\Lambda^{(l)}_{-}}d\Lambda'
\frac{2c}{c^2+(\Lambda^{(l)}-\Lambda')^2}\sigma^{(l)}(\Lambda')
\nonumber \\
&+&\int^{\Lambda^{(l-1)}_{+}}_{\Lambda^{(l-1)}_{-}}d\Lambda'
\frac{c}{(c/2)^2+(\Lambda^{(l)}-\Lambda')^2}
\sigma^{(l-1)}(\Lambda') \nonumber \\
&+&\int^{\Lambda^{(l+1)}_{+}}_{\Lambda^{(l+1)}_{-}}d\Lambda'
\frac{c}{(c/2)^2+(\Lambda^{(l)}-\Lambda')^2}
\sigma^{(l+1)}(\Lambda'), \quad 1\leq l\leq N-2
\label{eqn:density2}\\
2\pi\sigma^{(N-1)}(\Lambda^{(N-1)})&=&
\frac{1}{L}\frac{2c}{c^2+(\Lambda^{(l)})^2}
-\int^{\Lambda^{(N-1)}_{+}}_{\Lambda^{(N-1)}_{-}}d\Lambda'
\frac{2c}{c^2+(\Lambda^{(N-1)}-\Lambda')^2}\sigma^{(N-1)}
(\Lambda') \nonumber \\
&+&\int^{\Lambda^{(N-2)}_{+}}_{\Lambda^{(N-2)}_{-}}d\Lambda'
\frac{c}{(c/2)^2+(\Lambda^{(N-2)}-\Lambda')^2}
\sigma^{(N-2)}(\Lambda'), \label{eqn:density3}
\end{eqnarray}
where $\Lambda^{(0)}_{+}=-\Lambda^{(0)}_{-}=k_F^{(c)}$,
which is determined by the condition,
\begin{equation}
\int^{k_F^{(c)}}_{-k_F^{(c)}}dk \rho(k)=\frac{2M_0+1}{L}.
\end{equation}
In the absence of  magnetic fields, one can see that
$\Lambda^{(l)}_{+}=-\Lambda^{(l)}_{-}\rightarrow +\infty$
($1\leq l \leq N-1$), and the system recovers SU($N$)
spin symmetry. It is convenient to divide
the density functions into bulk and boundary parts,
\begin{equation}
\rho(k)\equiv \rho_{bulk}(k)+\frac{1}{L}\rho_{b}(k).
\end{equation}
In zero magnetic field, the coupled
integral equations for the density functions can be
reduced to simple ones for
$\rho_{bulk}$ and $\rho_{b}$ from eqs.(\ref{eqn:density1})
$\sim$(\ref{eqn:density3}) by using Fourier transformation.
The results are
\begin{equation}
2\pi\rho_{bulk}(k)=2+\int^{k_F^{(c)}}_{-k_F^{(c)}}
dk'G(k-k')\rho_{bulk}(k'),
\label{eqn:rho1}
\end{equation}
\begin{equation}
2\pi\rho_b(k)=\varphi_0'(k)+\varphi_L'(k)-
\frac{c}{(c/2)^2+k^2}+\int^{\infty}_{-\infty}dx e^{-ikx}
f(x)+\int^{k_F^{(c)}}_{-k_F^{(c)}}dk'G(k-k')\rho_{b}(k'),
\label{eqn:rho2}
\end{equation}
where the integral kernel is
\begin{equation}
G(k)=\int^{\infty}_{-\infty}dx e^{ikx}e^{-\frac{c}{2}\vert x\vert}
\frac{\sinh((N-1)cx/2)}{\sinh(N cx/2)},
\end{equation}
\begin{equation}
f(x)=\frac{e^{-c\vert x\vert}-e^{-\frac{c}{2}\vert x\vert}}
{2\sinh\frac{cx}{2}\sinh\frac{N cx}{2}}
(\cosh\frac{N cx}{2}+\cosh\frac{(N-1)cx}{2}
-\cosh cx-\cosh\frac{cx}{2})
+e^{-c\vert x\vert}\frac{\sinh\frac{cx}{2}}{\sinh\frac{N cx}{2}}.
\end{equation}
The total energy is given by
\begin{equation}
\frac{E}{L}=\int^{k_F^{(c)}}_{-k_F^{(c)}}dk\rho(k)k^2.
\label{eqn:ener}
\end{equation}
Eqs.(\ref{eqn:rho1}) $\sim$ (\ref{eqn:ener})
determine the energy spectrum of the model.
It should be noticed that in the thermodynamic limit, the
bulk properties of the present model is the same as
those for the periodic model\cite{suth},
which have been already studied in detail\cite{schlott}.
To examine critical properties, we thus need to
study the finite-size spectrum, which will be done below.

\section{Boundary Critical Properties }

In this section, we compute the energy spectrum for the
finite system, and discuss low-energy critical properties
using boundary conformal field theory.
We then obtain the surface
critical exponents of various correlation functions.

\subsection{finite-size spectrum and conformal properties}

Applying a standard technique in the Bethe ansatz method\cite{dw},
we now obtain the finite-size energy spectrum
from the basic equations derived in the previous section.
For this purpose, let us first express the total
energy, eq.(\ref{eqn:ener}), in terms of the dressed energies,
\begin{eqnarray}
\frac{E}{L}&=&\int^{k_F^{(c)}}_{-k_F^{(c)}}dk\biggl[\frac{1}{\pi}
+\frac{1}{2\pi L}(\varphi_0' (k)+\varphi_L' (k))
-\frac{1}{2\pi L}\frac{c}{(c/2)^2+k^2}\biggr]
\varepsilon^{(0)}(k) \nonumber \\
&&+\sum_{l=1}^{N-2}\int^{\Lambda^{(l)}_{+}}_{\Lambda^{(l)}_{-}}
d\Lambda\biggl[\frac{1}{2\pi L}\frac{2c}{c^2+\Lambda^2}
-\frac{1}{2\pi L}\frac{c}{(c/2)^2+\Lambda^2}\biggr]
\varepsilon^{(l)}(\Lambda)
\nonumber \\
&&+\int^{\Lambda^{(N-1)}_{+}}_{\Lambda^{(N-1)}_{-}}
d\Lambda\frac{1}{2\pi L}\frac{2c}{c^2+\Lambda^2}
\varepsilon^{(N-1)}(\Lambda),
\label{eqn:eps}
\end{eqnarray}
where the dressed energies $\varepsilon^{(l)}$
are determined by the following integral equations,
\begin{eqnarray}
\varepsilon^{(0)}(k)&=&k^2+
\int^{\Lambda^{(1)}_{+}}_{\Lambda^{(1)}_{-}}\frac{d\Lambda}{2\pi}
\frac{c}{(c/2)^2+(k-\Lambda)^2}\varepsilon^{(1)}(\Lambda) \\
\varepsilon^{(l)}(\Lambda)&=&
-\int^{\Lambda^{(l)}_{+}}_{\Lambda^{(l)}_{-}}
\frac{d\Lambda'}{2\pi}
\frac{2c}{c^2+(\Lambda^{(l)}-\Lambda')^2}
\varepsilon^{(l)}(\Lambda') \nonumber \\
&+&\int^{\Lambda^{(l-1)}_{+}}_{\Lambda^{(l-1)}_{-}}
\frac{d\Lambda'}{2\pi}
\frac{c}{(c/2)^2+(\Lambda^{(l)}-\Lambda')^2}
\varepsilon^{(l-1)}(\Lambda') \nonumber \\
&+&\int^{\Lambda^{(l+1)}_{+}}_{\Lambda^{(l+1)}_{-}}
\frac{d\Lambda'}{2\pi}
\frac{c}{(c/2)^2+(\Lambda^{(l)}-\Lambda')^2}
\varepsilon^{(l+1)}(\Lambda'), \quad 1\leq l\leq N-2 \\
\varepsilon^{(N-1)}(\Lambda)&=&
-\int^{\Lambda^{(N-1)}_{+}}_{\Lambda^{(N-1)}_{-}}d\Lambda'
\frac{2c}{c^2+(\Lambda^{(N-1)}-\Lambda')^2}\varepsilon^{(N-1)}
(\Lambda') \nonumber \\
&+&\int^{\Lambda^{(N-2)}_{+}}_{\Lambda^{(N-2)}_{-}}d\Lambda'
\frac{c}{(c/2)^2+(\Lambda^{(N-2)}-\Lambda')^2}
\varepsilon^{(N-2)}(\Lambda').
\end{eqnarray}
The equivalence of the
two expressions (\ref{eqn:ener}) and  (\ref{eqn:eps})
can be easily checked by directly comparing
them after formally solving the integral equations by the
iteration scheme. The formula (\ref{eqn:eps})
is particularly useful to compute the excitation
spectrum.

Let us start with the corrections to the ground state
energy. By directly applying the Euler-Maclaurin formula,
\begin{equation}
\frac{1}{N}\sum_{n=a}^b f\biggl(\frac{n}{N}\biggr)\sim
\int_{(a-1/2)/N}^{(b+1/2)/N}f(x)dx
+\frac{1}{24N^2}\biggl(f^{'}\biggl(\frac{a-1/2}{N}\biggr)
-f^{'}\biggl(\frac{b+1/2}{N}\biggr)\biggr), \label{eqn:em}
\end{equation}
to eq.(\ref{eqn:en}),
we easily find the finite-size corrections to the ground state
energy\cite{surface},
\begin{equation}
\Delta E_g=-\sum_{l=0}^{N-1}\frac{\pi v_l}{24L},
\label{eqn:eg}
\end{equation}
which is scaled by the velocities $v_l$ ($l=0, 1, \cdots, N-1$)
defined by
\begin{equation}
v_l=\frac{1}{2\pi \sigma^{(l)}(\Lambda^{(l)}_{+})}
\frac{\partial \varepsilon^{(l)}(\Lambda^{(l)}_{+})}
{\partial \Lambda^{(l)}},
\end{equation}
with $\sigma^{(0)}(\Lambda^{(0)})\equiv\rho(k)$.
By exploiting the finite-size scaling for open
boundaries\cite{bcn,cardy},
we can see from eq.(\ref{eqn:eg}) that
the Virasoro central charge for the charge sector
($l=0$) is given by $c=1$.  On the other hand,
all the velocities of the spin excitation takes the same
value ($l= 1, 2, \cdots, N-1$)
in the absence of magnetic fields,
and thus the central charge for the spin sector turns out to be
$c=N-1$, namely, the rank of SU($N$) Lie algebra.
We shall study conformal properties in more detail
by examining the excitation spectrum below.

Elementary excitations are classified into
two types, i.e. those for primary fields and
for descendant fields.
The former can be described by the excitations which
change the number of particles. They are computed by changing
the cut-off parameters in (\ref{eqn:eps})
as $\Lambda_{\pm}^{(l)} \rightarrow
\Lambda_{\pm}^{(l)} +\Delta \Lambda_{\pm}^{(l)}$.
On the other hand, the excitations for descendant fields are
simply given  by the particle-hole type excitations
with the fixed number of particles.
These manipulations are performed straightforwardly, and
we end up with the finite-size spectrum for the
excitation energy,
\begin{equation}
\frac{\Delta E}{L}=
\frac{\pi}{L}\biggl[\frac{1}{2}\Delta {\bf M}^{T}
(\hat{\xi}^{-1})^T V(\hat{\xi}^{-1})\Delta {\bf M}
+\sum_{l=0}^{N-1}v_l n^{(l)}_{+}\biggr],
\label{eqn:fss}
\end{equation}
with $V=\mbox{diag}(v_0, v_1, ... , v_{N-1})$, where $n^{(l)}_{+}$
are non-negative integers denoting particle-hole excitations.
Here $\hat{\xi}$ is the $N\times N$ matrix of the
so-called dressed charge
\cite{ikr}, whose components $\xi_{ij}\equiv
\xi_{ij}(\Lambda_{+}^{(i)})$
($0\leq i,j \leq N-1$)
are determined by the following integral equations,
\begin{eqnarray}
\xi_{0j}(k)&=&\delta_{0j}+
\int^{\Lambda^{(1)}_{+}}_{\Lambda^{(1)}_{-}}\frac{d\Lambda}{2\pi}
\frac{c}{(c/2)^2+(k-\Lambda)^2}\xi_{1j}(\Lambda) \\
\xi_{ij}(\Lambda^{(i)})&=&\delta_{ij}
-\int^{\Lambda^{(i)}_{+}}_{\Lambda^{(i)}_{-}}\frac{d\Lambda'}{2\pi}
\frac{2c}{c^2+(\Lambda^{(l)}-\Lambda')^2}
\xi_{ij}(\Lambda') \nonumber \\
&+&\int^{\Lambda^{(i-1)}_{+}}_{\Lambda^{(i-1)}_{-}}
\frac{d\Lambda'}{2\pi}
\frac{c}{(c/2)^2+(\Lambda^{(i)}-\Lambda')^2}
\xi_{i-1j}(\Lambda') \nonumber \\
&+&\int^{\Lambda^{(i+1)}_{+}}_{\Lambda^{(i+1)}_{-}}
\frac{d\Lambda'}{2\pi}
\frac{c}{(c/2)^2+(\Lambda^{(i)}-\Lambda')^2}\xi_{i+1j}(\Lambda'),
\quad 1\leq i\leq N-2 \\
\xi_{N-1j}(\Lambda^{(N-1)})&=&\delta_{N-1j}
-\int^{\Lambda^{(N-1)}_{+}}_{\Lambda^{(N-1)}_{-}}d\Lambda'
\frac{2c}{c^2+(\Lambda^{(N-1)}-\Lambda')^2}\xi_{N-1j}
(\Lambda') \nonumber \\
&+&\int^{\Lambda^{(N-2)}_{+}}_{\Lambda^{(N-2)}_{-}}d\Lambda'
\frac{c}{(c/2)^2+(\Lambda^{(N-2)}-\Lambda')^2}\xi_{N-2j}(\Lambda').
\end{eqnarray}
In eq.(\ref{eqn:fss}), the quantum numbers classifying  elementary
excitations are defined as
\begin{eqnarray}
&& \Delta M^{(l)} \equiv
\Delta M_h^{(l)}-\frac{n_b}{N}(N-l), \hskip 5mm
 1\leq l\leq N-1, \nonumber \\
&& \Delta  M^{(0)}=\Delta N_h-n_b.
\label{eqn:qn}
\end{eqnarray}
Here $n_b$ is the number of particles localized at boundaries,
which is given by
\begin{equation}
n_b=\frac{1}{2}\int^{k_F^{(c)}}_{-k_F^{(c)}}dk \rho_{b}(k),
\end{equation}
and $\Delta N_h$ is an integer specifying charge excitations,
whereas $\Delta M_h^{(l)}$'s are integers which label
$(N-1)$ kinds of spin excitations.

We can now read off conformal weights $\Delta_b$
from eq.(\ref{eqn:fss}), using finite-size scaling arguments
\cite{card}. Surface critical properties near the
boundary are determined by boundary scaling operators $\phi(t)$.
The critical exponent for correlation functions of a boundary
operator, $\langle \phi(t)\phi(0)\rangle \sim 1/t^x$, is given by
\begin{equation}
x=2\Delta_b=
\Delta {\bf M}^{T}
{\cal C}_{f}\Delta {\bf M}+2\sum_{l=0}^{N-1}n^{(l)}_{+},
\label{eqn:bcw}
\end{equation}
where the $N \times N$ matrix
${\cal C}_{f}=(\hat{\xi}^{-1})^T\hat{\xi}^{-1}$
is given in the absence of magnetic fields,
\begin{equation}
{\cal C}_{f}=
\left(
\matrix {\frac{1}{N K_{\rho}}+\frac{N-1}{N}   & -1  & \null  &
                 \smash{\lower1.7ex\hbox{\LARGE 0}} \cr
        -1     & 2  & \ddots & \null   \cr
        \null  & \ddots  &  \ddots     & -1      \cr
     \smash{\hbox{\LARGE 0}}   & \null   &  -1    & 2  \cr}
\right) .  \label{eqn:cf}
\end{equation}
Here $K_{\rho}\equiv \xi_{00}^2(k_F^{(c)})/N$ is the
dimensionless coupling constant for the charge sector
(so-called Tomonaga-Luttinger parameter), where
the dressed charge $\xi_{00}(k)$ is determined by the integral
equation,
\begin{equation}
\xi_{00}(k)=1+\int^{k_F^{(c)}}_{-k_F^{(c)}}dk G(k-k')\xi_{00}(k').
\label{eqn:dcxi}
\end{equation}
Note that the system is now regarded to be {\it chiral} due to
the presence of open boundaries, and quantum numbers carrying
currents do not appear in the conformal weights
(\ref{eqn:bcw}).
This implies that an effective theory of the present model is
given by the holomorphic piece of conformal field theory. From
eqs.(\ref{eqn:bcw}) and (\ref{eqn:cf}), we can see
that critical properties of the charge and spin sectors are
respectively described by
the U(1) Gaussian model with the central charge $c=1$
and the level-1 SU($N$) Wess-Zumino-Witten model with $c=N-1$.

The expression (\ref{eqn:bcw}) with (\ref{eqn:cf}) is
one of the main results
in this paper. We wish to emphasize that this formula is
quite general and is
applicable to many other SU($N$) quantum models with boundaries
such as the {\it t-J} model, the Hubbard model\cite{shu},  etc.


\subsection{surface critical exponents}

We are now ready to obtain the surface critical exponents of various
correlation functions.
As is seen from eq.(\ref{eqn:qn}), the effects of boundary
potentials are just to shift the quantum number as
$\Delta M^{(l)}\rightarrow\Delta M^{(l)}-\frac{n_b}{N}(N-l)$.
One readily notices that such an effect of
boundary potentials is equivalent to that
of  twisting boundary conditions, which does not
change the critical exponents in general.
Thus when we determine the critical
exponents from (\ref{eqn:bcw}), we should discard $n_b$-dependence
in $\Delta {\bf M}$ by redefining the quantum number as
$\Delta M^{(l)}-\frac{n_b}{N}(N-l)\rightarrow\Delta M^{(l)}$.
Therefore, for the long-time behavior of
the single-particle Green function,
$\langle c^{\dagger}_{\alpha}(t)c_{\alpha}(0)\rangle
\sim 1/t^{\eta}$, we obtain its critical exponent
by setting $\Delta N_h=1$, $\Delta M_h^{(l)}=0$ ($1\leq l \leq N-1$),
\begin{equation}
\eta=\frac{1}{N K_{\rho}}+\frac{N-1}{N}.
\label{eqn:eta1}
\end{equation}
Furthermore it is seen that the density-density correlation function
and the spin-spin correlation function show the following
asymptotic behavior, by taking $\Delta N_h=0$,
$\Delta M_h^{(l)}=0$ ($1\leq l \leq N-1$),
and $n^{(l)}_{+}=1$,
\begin{equation}
\langle O(t)O(0)\rangle \sim \mbox{constant}+\frac{1}{t^2}.
\label{eqn:dens1}
\end{equation}
Namely, this asymptotic behavior is  determined by descendants of
the primary field with $\Delta_b=0$, and hence the
critical exponent takes the canonical (integer) values.
The anomalous exponent appears only in the single-particle Green
function.  These characteristic properties are
inherent in the {\it chiral} Tomonaga-Luttinger liquid\cite{edge},
which is quite contrasted to ordinary  periodic systems as will
be seen in the next section.

We wish to stress that the formula (\ref{eqn:bcw})
for conformal dimensions possesses another important
information for boundary critical properties
related to the orthogonality catastrophe.
This is realized when one considers the problem
in which the boundary potentials are time-dependent\cite{al,fky2}.
For example, suppose that the boundary potentials are
suddenly turned on  at $t_0$. Then the long-time asymptotic
behavior of correlation functions
$\langle O(t_2)O(t_1)\rangle$ for $t_1\ll t_0\ll t_2$
should show quite different properties from
(\ref{eqn:eta1}) and (\ref{eqn:dens1}).
In this case, the phase shift caused by  boundary potentials
plays an essential role, and then $n_b$ cannot be
ignored by redefining the quantum numbers. Therefore,
retaining $n_b$ in (\ref{eqn:bcw}),
we get anomalous exponents for various correlation
functions.  We show several examples below.

\noindent
(i) single-particle Green function:
$\langle c^{\dagger}_{\alpha}(t)c_{\alpha}(0)\rangle
\sim 1/t^{\eta}$. Putting $\Delta N_h=1$,
$\Delta M_h^{(l)}=0$ ($1\leq l \leq N-1$), we have
\begin{equation}
\eta=\frac{1}{N K_{\rho}}(1-n_b)^2+\frac{N-1}{N}.
\label{eqn:eta2}
\end{equation}

\noindent
(ii) density-density correlation function:
$\langle n(t)n(0)\rangle \sim 1/t^{\alpha_c}$, ($n=\sum_{\alpha}
c^{\dagger}_{\alpha}c_{\alpha}$).
Taking $\Delta N_h=0$, $\Delta M_h^{(l)}=0$ ($1\leq l \leq N-1$),
we then obtain
\begin{equation}
\alpha_c=\frac{n_b^2}{NK_{\rho}}.
\label{eqn:char}
\end{equation}

\noindent
(iii) spin-spin correlation function:
$\langle S^a(t)S^a(0)\rangle \sim 1/t^{\alpha_s}$,
($S^a=\sum_{\alpha,\beta}
c^{\dagger}_{\alpha}\tau^a_{\alpha\beta}c_{\beta}$
with $\tau^a_{\alpha\beta}$ being
a fundamental representation of SU($N$)). Putting
$\Delta N_h=0$, $\Delta M_h^{(l)}=0$ ($1\leq l \leq N-1$),
we have
\begin{equation}
\alpha_s=\frac{n_b^2}{NK_{\rho}}.
\label{eqn:spin}
\end{equation}

Thus the surface critical exponents depend not only on
the Tomonaga-Luttinger parameter $K_{\rho}$
but also on the fractional number $n_b$ of localized particles
at the boundary.  We will see in the next section that
these critical properties have a deep connection
to the X-ray edge problem
in 1D chiral systems.

We can evaluate the above exponents easily
in some limiting cases. In the case of noninteracting
electrons ($c=0$), we see
$K_{\rho}=1$ from (\ref{eqn:dcxi}),
and the critical exponents (\ref{eqn:eta1}), (\ref{eqn:dens1})
(\ref{eqn:eta2}), (\ref{eqn:char}) and (\ref{eqn:spin})
are all reduced to those obtained previously
for the single-impurity SU($N$) Anderson model with infinitely
strong Coulomb interaction\cite{fky2}. On the other hand,
in the strong-coupling limit $c\rightarrow +\infty$ or low-density
limit $k_F^{(c)} \rightarrow 0$, it is easily
seen from eq.(\ref{eqn:dcxi})
that $K_{\rho}=1/N$, and so the exponent of the single-particle
Green function eq.(\ref{eqn:eta1}) is $\eta=2-1/N$.
In general, $K_{\rho}$ may change in the range $[1, 2-1/N]$
according to the interaction strength and the density
of particles.

\subsection{comparison with the periodic model}

To conclude this section, we compare the above results
with bulk critical exponents under periodic boundary conditions.
The finite-size spectrum for the SU($N$) model in the
periodic boundary conditions is given by\cite{ikr}
\begin{equation}
\frac{E}{L}=\frac{2\pi}{L}\biggl(\frac{1}{4}\Delta {\bf M}^{T}
(\hat{\xi}^{-1})^T V(\hat{\xi}^{-1})\Delta {\bf M}
+{\bf D}^T\hat{\xi}V\hat{\xi}^T {\bf D}+
\sum_{l=0}^{N-1}v_l (n^{(l)}_{+}+n^{(l)}_{-})
\biggr),
\end{equation}
where the newly introduced
quantum numbers, ${\bf D}^T=(D_0, D_1, ..., D_{N-1})$,
are given by
\begin{equation}
D_0=\frac{\Delta M^{(0)}+\Delta M^{(1)}}{2}, \qquad (\mbox{mod } 1),
\end{equation}
\begin{equation}
D_l=\frac{\Delta M^{(l-1)}+\Delta M^{(l+1)}}{2},
\quad (\mbox{mod } 1), \quad l=1,...,N-1,
\end{equation}
with $\Delta M^{(N)}\equiv 0$.
These quantum numbers carry the large momentum transfer\cite{ikr}.

Using finite-size scaling arguments
for periodic systems\cite{card},
we have the conformal dimensions of scaling operators,
\begin{equation}
\Delta_{+}+\Delta_{-}=\frac{1}{4}\Delta {\bf M}^{T}
{\cal C}_{f}\Delta {\bf M}+{\bf D}^T{\cal C}_f^{-1}{\bf D}
+\sum_{l=0}^{N-1}(n^{(l)}_{+}+n^{(l)}_{-}).
\end{equation}
\null From this formula, we obtain critical exponents.
We list several examples in the following.

\noindent
(i) The single-particle Green function:
\begin{equation}
\eta=\frac{1}{2N}\biggl(\frac{1}{K_{\rho}}+K_{\rho}\biggr)
+\frac{N-1}{N}.
\end{equation}
In the strong-coupling limit or low-density limit,
this reduces to \cite{kawa,frahm}
\begin{equation}
\eta=\frac{3}{2}-\frac{1}{N}+\frac{1}{2N^2}.
\end{equation}

\noindent
(ii) The $2Nk_F$-oscillating term ($k_F$: Fermi momentum)
of the density-density correlation function:
\begin{equation}
\alpha_c=2NK_{\rho}.
\end{equation}
We have $\alpha_c=2$ in
the strong-coupling limit or low-density limit.

\noindent
(iii) The $2k_F$--oscillating term of the
spin-spin correlation function:
\begin{equation}
\alpha_s=2\frac{K_{\rho}+N-1}{N},
\end{equation}
which reduces to
\begin{equation}
\alpha_s=2\biggl(1-\frac{1}{N}+\frac{1}{N^2}\biggr),
\end{equation}
in the strong-coupling limit or low-density limit\cite{kawa}

Comparing these results with the previous ones,
we can see that the bulk critical exponents are quite
different from the boundary ones, though bulk thermodynamic
properties
should exhibit the same behavior in both cases.

\section{Application to the edge state of FQHE}

In this section, we briefly discuss a possible  application of
the results obtained in the previous section
to  the edge state of the FQHE as
a typical example of 1D chiral fermion systems\cite{edge}.
We first note that if an appropriate
value is chosen for the Tomonaga-Luttinger parameter $K_\rho$,
the above SU($N$) model with open boundaries
can give an effective theory for the edge state of
a certain hierarchy in the FQHE.
Namely, the edge state of the fractional quantum Hall effect
with filling $\nu=N/(Nm+1)$ ($m$ even)
can be modeled by the above open system
by choosing $K_{\rho}$ as $K_{\rho}=\nu/N$\cite{fky2}.
In fact, one can easily check that by this choice of
$K_{\rho}$, the formulas (\ref{eqn:eta1}) and (\ref{eqn:dens1})
exactly reproduce the critical exponents for the
chiral liquids proposed for the edge states \cite{edge}.

We shall focus on the X-ray photoemission (absorption)
problem in edge states.
The Fermi-edge singularity problem in 1D electron systems
has attracted current interest\cite{lut1,lut2,lut3}.
In the edge state of the FQHE, electrons
move only in one direction and the backward
scattering due to impurities is irrelevant, and hence the system
can be treated  as a chiral system. In experiments of the
X-ray photoemission or absorption, the core hole
may be suddenly created, resulting in  a
problem with time-dependent boundary conditions\cite{fky2,al}.
That is, for $t<t_0$ the boundary is free, and at $t=t_0$ the
boundary potential suddenly switches on.
Then bulk electrons show critical low-energy behavior
inherent in the orthogonality catastrophe.

In the X-ray absorption problem, one electron is excited
in the final state.
Thus putting $\Delta N_h=1$, $\Delta M^{(l)}_h=0$
($1\leq l\leq N-1$), we have
the critical exponent for the X-ray absorption in this system,
\begin{equation}
\alpha_{ab}=\frac{1}{\nu}\biggl(1-\frac{N\delta}{\pi}\biggr)^2
+\frac{N-1}{N},
\end{equation}
for the filling  $\nu=N/(Nm+1)$ with even $m$,
where $\delta$ is the phase shift
caused by the localized electrons which screen
the core hole potential.
On the other hand, the critical exponent for the
photoemission is obtained by
taking $\Delta N_h=0$ and $\Delta M_h^{(l)}=0$
in eq.(\ref{eqn:bcw}),
because one hole carrying neither charge nor spin is generated
in the final state,
\begin{equation}
\alpha_{ph}=\frac{N^2}{\nu}\biggl(\frac{\delta}{\pi}\biggr)^2.
\end{equation}
We expect that such anomalous exponents may be
observed in the X-ray photoemission or absorption
experiments  for the edge state of the FQHE
of the filling $\nu=N/(Nm+1)$ with even $m$.

\section{Summary}

We have studied exact boundary critical properties of the SU($N$)
interacting fermion model with open boundaries
by using the Bethe ansatz solution and boundary conformal field
theory. It has been shown that the
surface exponents which govern the critical behavior
near the boundary depend on the two continuously
varying quantities, i.e. the dimensionless Tomonaga-Luttinger
parameter $K_\rho$  and the fractional number of localized
electrons at the boundary.
We have also discussed a possible  application of the results
to the Fermi-edge singularity problem in the edge state of the FQHE
as a typical example of 1D chiral systems.
The exact exponents for the X-ray absorption and photoemission
have been derived.

\acknowledgments{}
We wish to express our sincere thanks to T. Fukui,
T. Yamamoto and S.-K. Yang for valuable discussions.
This work was partly supported by a Grant-in-Aid from the Ministry
of Education, Science and Culture.

\newpage

\newpage


\begin{references}
\bibitem{surface} In eq.(\ref{eqn:eg}), we omit the correction
term of the zeroth order in $1/L$ which corresponds to
the non-universal surface energy of the boundaries, since it is
irrelevant to the following discussion
about the critical properties.
\end{references}
\end{document}